\documentclass[epj]{svjour}
\usepackage{graphics}
\usepackage{slashed}
\usepackage{graphicx}
\usepackage{subfigure}
\usepackage{dcolumn}
\usepackage{hyperref}      
\usepackage{bm}
\usepackage{epsfig}
\usepackage{setspace}

\newcommand{\eg}{{\em e.g.}}

\newcommand{\eqref}[1]{Eq.~(\ref{#1})}

\newcommand{\beq}{\begin{equation}}
\newcommand{\eeq}{\end{equation}}
\newcommand{\beqa}{\begin{eqnarray}}
\newcommand{\eeqa}{\end{eqnarray}}
\newcommand{\nn}{\nonumber}

\newcommand{\slep}{{\tilde{l}}}
\newcommand{\ES}{E_{{\rm shift}}}

\begin{document}
\title{Flavor and LHC Searches for New Physics}
\author{Yael Shadmi\inst{1}
}                     
\offprints{}          
\institute{
Physics Department, Technion---Israel Institute of 
Technology, Haifa 32000, Israel}
\date{Received: date / Revised version: date}
%
\abstract{
Uncovering the physics of electroweak symmetry breaking (EWSB) is
the raison-d'etre of the LHC. Flavor questions, it would seem,
are of minor relevance for this quest, apart from their role in
constraining the possible structure of EWSB physics.
In this short review article, we outline, using flavor-dependent sleptons
as an example, how flavor can affect both  searches for supersymmetry,
and future measurements aimed at understanding the
nature of any new discoveries.
If the production cross-sections for supersymmetry
are relatively low, as indicated by the fact that it has not
revealed itself yet in standard searches, 
the usual assumptions about the superpartner spectra need re-thinking.
Furthermore,
one must consider
more intricate searches, such as lepton-based searches, 
which could be susceptible to flavor effects.
We start by reviewing the flavor structure of 
existing frameworks for mediating supersymmetry breaking, emphasizing 
flavor-dependent models proposed recently.
We use the kinematic endpoints of invariant mass distributions to
demonstrate how flavor dependence can impact both searches for supersymmetry
and the Inverse Problem.
We also discuss methods for measuring small-mass splittings and
mixings at the LHC, both in models with a neutralino LSP and in
models with a charged slepton (N)LSP.
\PACS{      {12.60.Jv}{Supersymmetric models} \and
       {14.80.Ly}{Supersymmetric partners of known particles} \and
 {11.30.Hv}{Flavor symmetries}
     } 
} 
\maketitle
%
\section{Introduction}
\label{intro}
The subject of this review is the interplay between flavor questions
and LHC searches for the origin of electroweak symmetry breaking.
The latter is the most pressing problem of the standard model (SM).
With just the observed SM particles, this theory fails
around the TeV scale, because it predicts $W$ and $Z$ scattering
cross-sections above the unitarity bounds. 
If the $W$ and $Z$ cross section is unitarized by a Higgs scalar,
one is still faced with the hierarchy, or fine-tuning, problem.
This is of course the raison-d'etre of the LHC, and the main
argument for new particles at or below the TeV.

Flavor\footnote{The term ``flavor'' is somewhat vague,
and often redundant, since it can usually be replaced by ``generation.'' 
The matter fields of the SM carry both gauge indices and generation
indices. The latter are often called
flavor indices. Flavor-dependence simply means generation-dependence,
and flavor parameters refer to the generation-dependent
parameters of the SM.}, 
too, refers to one of the open issues in the 
SM. The SM flavor puzzle is the fact that the intricate generation structure
of the SM has no fundamental explanation. Both the fermion masses
and their couplings to the $W$ bosons involve many small and hierarchical
dimensionless parameters with no explanation in the 
SM.
This puzzle is really a question of aesthetics. 
Not only is there no fine-tuning involved, but the small parameters are
technically natural. There is certainly no inherent inconsistency
in the SM with its unexplained generation structure.
Still, it seems to suggest an underlying flavor theory,
one that would naturally generate the observed masses and mixings. 

There are several ways in which flavor physics, and 
the new physics (NP) motivated by electroweak symmetry breaking
(which we will refer to simply as NP), are related.
We list these here, and will elaborate on each of them
in the next sections.

{\bf (i) Bounds on flavor violation constrain the NP.}
The NP could
exhibit a non-trivial generation structure. 
At the very least, if the new states couple directly to
the SM fermions, the couplings could apriori be generation dependent.
Furthermore, the new states could also appear in three
copies, corresponding to the three generations of the SM.
This is necessarily the case if this NP is the consequence of
enlarging the space-time symmetry of the SM, as in supersymmetric
and extra-dimension models.
The SM matter fields then have new partners that come
in three copies,
and the masses and couplings of the new states are  matrices, or
more generally higher-rank tensors,
in generation space.
If these matrices are arbitrary, and the new states are around the TeV,
the stringent bounds on flavor violation would be grossly 
violated\footnote{For a review of EDM bounds on CP-violating phases,
see the article by B.~Batell in this chapter.}.
This greatly constrains the structure of the NP, and has led
to the spreading belief that the NP must be 
Minimally Flavor Violating (MFV),
that is, the only source of generation dependence of the NP
can be the SM Yukawas~\cite{D'Ambrosio:2002ex}.

That this assumption needs rethinking is  clear.
In the case of supersymmetry for example, the fact that it has not shown
up yet at the LHC may hint at possibilities that are markedly
different from MFV scenarios. 
The most drastic possibility, if supersymmetry exists, 
is that all sfermions, or at least all squarks, are heavy,
as in~\cite{Feng:1999zg,Giudice:2004tc,ArkaniHamed:2004fb}, 
and beyond the LHC reach. 
This may imply a certain amount of fine-tuning in the Higgs sector,
and would altogether eliminate flavor 
constraints for scalar masses of~$10^4$~TeV~(see, \eg, \cite{Isidori:2010kg}). 
A more moderate possibility is that only the first- and second-generation 
squarks are heavy, as in More Minimal Supersymmetry 
Breaking~\cite{Cohen:1996vb,Dvali:1996rj}.

At this stage, however, given the (relatively) low energy of the LHC,
the small dataset analyzed so far (1~fb$^{-1}$), and the
simplifying assumptions made in many of the analyses,
it is still possible that more mundane supersymmetric models are 
within LHC reach. Even in this case, as was emphasized in recent
years (see \eg, \cite{Feng:2007ke,Nomura:2008gg,Kribs:2007ac}), 
there is room for generation-dependent superpartner spectra
that are nonetheless consistent with low-energy bounds,
and there are natural models that predict such spectra.

{\bf (ii) Flavor impacts NP search strategies.}
The flavor structure of the NP obviously determines its
LHC signatures. 
It is important to bear in mind that most studies of NP at the LHC
assume a flavor-blind, or MFV spectrum. 
In the busy LHC environment, however, if we don't specifically
look for something, it can be easily missed.
That More Minimal Supersymmetry
may be missed by traditional search techniques
is clear (see for example the recent analyses 
of~\cite{Baer:2010ny,Brust:2011tb}) 
but the same could hold for more ``bread-and-butter''
models, especially if only a subset of the superpartners
are light enough to be produced.

 {\bf (iii) Flavor can exacerbate the Inverse Problem.}
Even if NP is discovered at the LHC, understanding the
nature of this NP, and whether and how it is related to EWSB,
would be far from trivial (see, \eg,~\cite{Feng:1995zd,ArkaniHamed:2005px}). 
This task would be especially
hard if only a few new particles are discovered.
In particular, it is well known that different solutions
to the hierarchy problem, such as supersymmetry and UED models,
have similar LHC signatures, and would therefore be hard to tell
apart. This has been called the 
``Inverse Problem''~\cite{ArkaniHamed:2005px}.
This problem can be even more challenging if the NP is
generation dependent.
We will see an example of this kind in~section~\ref{inverse}.

{\bf (iv) How well can we measure the NP flavor parameters?}
If NP is discovered at the LHC, one would like to understand
its flavor structure. 
Are the new states single states, with universal couplings to different
SM generations? Are they single states with generation-dependent
couplings? Are there three copies of new states with different or equal
masses and with different couplings to the SM generations?
It is therefore important to devise methods for measuring the
flavor parameters of the NP. 
A particularly plausible possibility, given the stringent bounds
on flavor violation involving the first and second generations,
is that some of the new states are almost degenerate, especially
if these states are far from pure flavor states.
Such a scenario would be hard to distinguish from having a single state
which couples to different SM generations, or from
having two degenerate pure flavor states.

{\bf (v) The NP flavor parameters would provide information about the
underlying NP theory.}
Thus for example, they could tell us about the mediation mechanism
of supersymmetry breaking.

 {\bf (vi) The NP flavor parameters may provide information about the
underlying flavor theory.}
If the SM generation structure is explained by some flavor theory,
this flavor theory may control the generation structure
of the NP. The NP  flavor parameters will then provide
new handles on the flavor theory, whether this theory involves a
flavor symmetry, new strong dynamics, or the wave-functions
of the matter fields in an extra-dimension.

There is a huge  literature on flavor, and it's impossible
to coherently review it within the limited length of this review.
We will therefore focus on a very specific example, namely the
sleptons of supersymmetry, in order to demonstrate the
inter-relations listed above. 
There are several reasons why sleptons are a useful example.
First, as opposed to quarks, (especially the light quarks), 
the flavor of a lepton--whether it is an electron, a muon, 
and to a lesser extent, a tau---can be determined very precisely in collider
detectors.
Second, many frameworks for mediating supersymmetry breaking
predict sleptons at the bottom
of the spectrum, below colored particles, so that they typically
appear at the end of decay chains and will therefore be
easier to reconstruct.
For the same reason, if colored particles are above LHC reach,
we may have to content ourselves with sleptons and electroweak gauginos
only. These have much smaller production cross sections,
so may still be hiding in LHC data.
Finally, the lepton inter-generation mixings, measured in neutrino
oscillations,  are much larger than the quark inter-generation mixings.
These mixings could feed into the slepton inter-generational
mixings, resulting in observable lepton flavor-changing processes
at the LHC~(see, \eg, \cite{Feng:1999wt}). 
It should also be noted that much of the above reasoning in favor of sleptons
can be repeated for different lepton partners, such as KK leptons.

\section{The mediation of supersymmetry breaking in light of 
flavor-constraints} \label{constraints}
Supersymmetry would not have much theoretical appeal if we didn't have
plausible theoretical frameworks for generating the 100 or so 
soft supersymmetry-breaking terms from a few fundamental parameters.
Flavor is an important guiding principle in devising such frameworks,
since weak-scale superpartners generically 
give rise to large loop contributions to
flavor changing processes, such as $\mu\to e\gamma$ or $K^0-\bar K^0$ mixing.
 
The experimental constraints on flavor violation are usually derived
by working in the interaction basis, in which the sfermion-fermion-gaugino
coupling is a diagonal matrix in generation space,
using the mass-insertion approximation~\cite{Gabbiani:1996hi},
\beq
\delta^{MN}_{ij}\equiv\frac{\Delta \tilde M^{2,MN}_{i\neq j}}{ \tilde m^2} \ll1\,.
\eeq
Here $M,N=L,R$, with $L,R$ denoting the sfermion partners of left-handed
or right handed-fermions, $\tilde M^{2,MN}$ is the mass squared matrix for 
the relevant sfermions,
and $\tilde m^2$ is the typical sfermion mass squared.
The bounded quantities are then
the off-diagonal elements in the sfermion mass-squared
matrices, which provide useful information for  top-down model building. 
When discussing the collider signatures of superpartners, however, it is
more natural to work in terms of the physical observables---the
sfermion masses and mixings. 
These are found by diagonalizing the $6\times6$ mass matrix
for the three $L$-sfermions and three $R$-sfermions.
The gaugino coupling to a (mass-eigenstate) fermion of generation $i$
and sfermion $J$ is then proportional to some mixing matrix
$K_{iJ}$, 
where $J=1,\ldots,6$ runs over the six mass eigenstates.

There are several possible approaches
to suppressing the superpartner loop contributions
to FV processes.
Essentially, these follow from the different limits in which the
FV contributions vanish.
Since we would first like to enumerate different approaches to mediating
supersymmetry breaking, it is convenient to phrase the discussion 
in terms of the theoretical input parameters $\tilde M^{2,MN}$.
Often, these contain a few sources of FV and
one or two of the approaches below must be combined. 
Below we list features of the soft terms that lead to reduced FV,
starting with the scalars and continuing with the 
gauginos\footnote{To better understand these points, 
one should examine the structure
of FV diagrams. See for example
eqn.~20 of~\cite{Feng:2001sq}.}:
\begin{enumerate}
\item Heavy Scalars (often called decoupling): 
Some diagonal elements of $\tilde M^{2,MM}$ are much larger than
a TeV$^2$, so that the corresponding scalars are heavy.
\item Alignment: The matrices $\tilde M^{2,MN}$ 
are approximately diagonal in the fermion mass basis.  
\item Universality: $\tilde M^{2,MN}\propto 1_{3\times 3}$.
\item Small $A$ terms. Penguin-type FV processes 
like $\mu\to e\gamma$ require a helicity flip on the fermion line. 
If they don't involve Higgsinos, the diagrams therefore
require an insertion of  $\delta^{LR}$. 
These can be dangerous even if they are not flavor-violating,
since multiple insertions involving a flavor-diagonal 
$\delta^{LR}$ can dominate over a single insertion.
\item Heavy gauginos: Since the loop diagrams
contain both virtual spin-0 sfermions {\sl and} spin-1/2 
gauginos or Higgsinos it suffices to take {\sl either} the sfermions
{\sl or} the spin-1/2 superpartners to be heavy.
\item Gauginos are Dirac fermions: 
In this case, Penguin type diagrams with 
a helicity flip on the gaugino  line are absent.
\end{enumerate}
We can now classify existing models and frameworks 
according to which of these features they employ\footnote{Some
of these are actual models, with a well-defined mechanism
for generating the soft terms and controlling their structure.
Others are simply ansatze, which may have some concrete realization(s).}.

Heavy superpartners occur in different scenarios.
Indeed, 
the null results of LHC searches for supersymmetry could
indicate that some, or all, scalars are heavy.
Models in which all scalars are beyond LHC reach are not
relevant for this review. 
Dirac gauginos~\cite{Fox:2002bu}, as in $R$-symmetric 
models~\cite{Kribs:2007ac}, 
can be heavier than the scalar superpartners
since their effect on the Higgs mass is relatively small.
$R$ symmetric models therefore
allow\footnote{These models have additional 
features 
that contribute to the suppression of FV processes, including the absence of
$A$ terms and the Dirac nature of the gauginos (see items 5, 6 above).} 
for generation-dependent slepton and
squark masses~\cite{Kribs:2007ac,Fok:2010vk}.
Similarly, the first-and second-generation  squarks can be
heavy with only mild tuning of the Higgs mass~\cite{Dvali:1996rj,Cohen:1996vb}.
This pattern is sometimes called an ``Inverted Hierarchy''
or ``More Minimal Supersymmetry''.
The flavor constraints and LHC signatures of this scenario
were recently studied in~\cite{Baer:2010ny,Brust:2011tb}).

Universality is achieved if the soft terms are determined
solely by the gauge charges of the scalars.
This is the case for the soft terms of 
Gauge Mediated Supersymmetry Breaking (GMSB)~\cite{Dine:1994vc,Dine:1995ag}
or Gaugino Mediated Supersymmetry 
Breaking (gMSB)~\cite{Chacko:1999mi,Kaplan:1999ac}
at the scale at which supersymmetry breaking is mediated to the MSSM.
The $A$ terms at the mediation scale are zero in these models.
 ``Minimal SUGRA'' (mSUGRA)
 imposes universality at the mediation scale as
an ansatz, and actually assumes that all scalar masses are the same at 
this scale\footnote{In gravity mediated models, without additional ingredients,
the scalar mass matrices could be arbitrary,
so that bounds on FV are not satisfied.}.
In Anomaly Mediated Supersymmetry Breaking (AMSB)~\cite{Randall:1998uk},
the soft terms at the mediation scale are only approximately universal.
They are determined by the anomalous
dimensions of the MSSM fields, and therefore involve both the
gauge couplings and the Yukawa couplings. 

In all these models, the soft terms acquire some Yukawa dependence
due to the RG evolution  to the weak scale.
These models are then
``Minimally Flavor Violating'' (MFV):
the only source of generation dependence is the  
SM Yukawa couplings~\cite{D'Ambrosio:2002ex}. 
The size of the Yukawa dependence is affected by the amount of running,
and is therefore the smallest
in GMSB models with a low mediation scale. 

Alignment can be achieved when the underlying theory
responsible for generating the SM fermion mass matrices also
controls the soft terms~\cite{Nir:1993mx}. 
Known examples include models with flavor symmetries~\cite{Nir:1993mx}, 
(or Froggatt-Nielsen symmetries~\cite{Froggatt:1978nt}), 
and theories in which the SM fields develop 
large anomalous dimensions through their couplings to a near-conformal
theory~\cite{Nelson:2001mq}.

Small $A$ terms arise naturally in some of the frameworks mentioned
above such as GMSB and gMSB. They are also a feature of R-symmetric
models, since they break R-symmetry. Other mechanisms for
suppressing these include the Higgsophobic supersymmetry 
breaking of~\cite{Nomura:2007ap,Nomura:2008gg}, which relies on a fifth dimension to separate the supersymmetry-breaking fields from the Higgses.

As mentioned above, for the purposes of this review, the parameters
of interest are the physical parameters in the slepton mass basis, 
namely, the maximal mass splittings and mixings 
consistent with FV bounds.
Universality and small $A$ terms result in mass degeneracy.
Alignment results in small mixings.

As we saw above, models exhibiting universality are MFV.
Since the only order one Yukawa coupling is the top
Yukawa (except at large $\tan\beta$, for which the bottom and
tau Yukawas can be significant) 
the charged matter sector of the low-energy 
supersymmetric models have an approximate 
SU(2)$_{U_L}\times$SU(2)$_{U_R}\times$
SU(3)$_{D_L}\times$SU(3)$_{D_R}\times$SU(3)$_{\rm lepton}$ symmetry.  
To see interesting flavor effects at the LHC, we must therefore
go beyond MFV. The models must involve
some of the remaining ingredients listed above: alignment,
heavy scalars or gauginos, or Dirac gauginos.

To be concrete, let us focus from now on on the sleptons.
We can estimate the sizes of the slepton mass matrices 
using their transformation properties under the
 ${\rm SU}(3)_L\times{\rm SU}(3)_e$ flavor symmetry, broken
by the charged lepton Yukawa $Y_L$ which transforms as a bi-triplet
of this 
symmetry\footnote{SU(3)$_L$ is further broken by the neutrino 
masses.  Here, for simplicity, we will neglect the 
spurions associated with the neutrinos. 
This is justified if the seesaw scale is higher than the mediation scale.
We can then take the charged lepton Yukawa to be a diagonal
matrix without loss of generality.}. 
To leading order in $Y_L$, the slepton soft terms are then
\beqa
M^2_L &=& \tilde m^2_L \left(1_{3\times3} + Y_L Y_L^\dagger + \ldots\right)\,,\nn\\ 
M^2_e &=& \tilde m^2_e \left(1_{3\times3} +  Y_L^\dagger Y_L + \ldots\right)\,, \\ 
A_{LR} &=& \tilde m_{LR}\left(Y_L + \ldots\right)\,, \nn
\eeqa
where $\tilde m^2_L$, $\tilde m^2_e$ and $\tilde m_{LR}$ are numbers.
We then find that, for LHC purposes, the spectrum is almost generation
blind. The only potentially observable
effect is the stau splitting from the other sleptons, 
with 
${\Delta \tilde m_{I3}}/{\tilde m}\sim 10^{-4}\tan^2\beta$ (with $I=1,2$)
which for $\tan\beta\sim m_t/m_b$ is roughly 10$\%$.
In order to have either slepton mixings, or selectron-smuon mass
splittings, we must go beyond MFV.

We now give a few representative examples.\\
$\bullet$ Flavor constraints satisfied by a combination of universality
and alignment:
Some examples of this type are based
a dominant universal contribution
from GMSB~\cite{Feng:2007ke,Shadmi:2011hs}, or 
AMSB~\cite{Gross:2011gj}\footnote{As mentioned above, AMSB is not universal 
but MFV.}, with a sub-dominant non-universal contribution
controlled by the flavor symmetry responsible
for the structure of the SM fermion masses.
This can be realized purely in GMSB models~\cite{Shadmi:2011hs}, 
with the non-universal
contribution coming from superpotential matter-messenger couplings,
or in high-scale GMSB~\cite{Feng:2007ke} models which always have sub-dominant
but non-negligible gravity-mediated contributions to the soft masses.
Such models typically have small LR mixings.
The right handed sleptons can have order-one relative mass splittings,
with small mixings: $1-2$ mixings at the percent level 
and $2-3$ mixings at the ten percent level\footnote{These small mixings
are a generic result of the flavor symmetry.}.
The left handed sleptons can have relative mass-splittings of order
one with negligible mixings, or, at the other extreme, order-one mixings
with relative mass splittings of order 10$^{-3}$ (such splittings
will probably be below LHC sensitivity).
Between these two extremes,
there are models in which both the relative mass splittings 
and the mixings are a few percent.

$\bullet$ $R$-symmetric Models\cite{Kribs:2007ac}. 
Here FV is suppressed by a combination
of having Dirac gauginos 
(some of which can additionally be heavy---see item 4 above), and the
absence of $A$ terms.
Ref.~\cite{Fok:2010vk} performed a detailed analysis of the predictions 
of a class of $R$ symmetric models
to $e-\mu$ FV processes, and showed that even for large selectron-smuon
mass splittings of 50$\%$, 10$\%$ mixings are generically allowed.
In small wedges of the parameter space, even O(1) mixings are possible.

$\bullet$ Supersymmetric 5d models. The SM flavor parameters may be the result
of different overlaps of the 5d fermion wave-functions with
the Higgs wave function, in either flat or warped 5d 
models~\cite{ArkaniHamed:1999dc,Gherghetta:2000qt}.
Since the hierarchy problem is solved by supersymmetry,
the 5th dimension can be very small in this case, thus for example
it can be of GUT-scale size.
The 5d locations of the matter fields will generically 
affect the soft terms~\cite{Kaplan:2000av,Kaplan:2001ga}.
As mentioned above, in Higgsophobic models, the Higgs fields
and supersymmetry breaking fields are localized at different points
along an extra dimension. In this case, $A$ terms are adequately
suppressed. The resulting R-sleptons can then exhibit relative mass splittings
of ${\cal O}(10^{-2})$ between the selectron and smuon, with
order one mass splittings from the stau~\cite{Nomura:2007ap,Nomura:2008gg}.
A detailed analysis of flavor in 5D models
recently appeared in~\cite{Brummer:2011cp}. In these models, the Higgses and the
supersymmetry breaking sector are taken to be on the same 4d brane.
The ${\cal O}(1)$ 5d mass-parameters of the matter fields (which determine
the wave-functions of these fields) are chosen so that viable
fermion masses are obtained. Flavor constraints then imply several
allowed superpartner spectra, with all colored superpartners 
above LHC reach. The lightest superpartners are the $R$-handed sleptons
(often the selectron or smuon), with masses that are rather
large, around 500-600~GeV and relative mass splittings of roughly
20\%, and mixings of order 10\%.

To summarize, we have examples of viable models that predict
potentially observable generation-dependent slepton spectra at
the LHC, with mass splittings of the selectron and smuon
of up to tens (or even hundreds, in $R$-symmetric models) of GeV,
and with slepton mixings up to order one. Some examples
can exhibit selectron and smuon splittings above a few GeV
with non-negligible mixings.
 
\section{Flavor impacts NP search strategies.}\label{search}
The basic supersymmetry searches rely on jets plus missing $E_T$
and are therefore truly blind to 
slepton flavor\footnote{Flavor-dependent squark spectra and their
impact on LHC searches were discussed for example 
in~\cite{Hurth:2011zy,Hurth:2003th,Giudice:2011ak}.}.
Searches involving leptons are especially important given the fact that colored
superpartners have so far not revealed themselves at the LHC.
If squarks and gluinos are beyond LHC reach, and only neutralinos, charginos
and sleptons can be produced, lepton-based searches would 
be essential (see, \eg,~\cite{Eckel:2011pw}).
Lepton searches are important even if
colored superpartners are within reach, but are very heavy.
In this case, the electroweak cross sections for producing
neutralinos, charginos, and sleptons can become comparable
to strong SUSY production.
Relying on missing $E_T$ is then problematic, since
the masses of the pair produced non-colored particles is
not that much higher than the top, $W$ and $Z$ masses.
The missing $E_T$ carried by a neutralino $\tilde \chi_1^0$
that is  produced from cascade decays of heavier charginos or neutralinos, 
is likely to be smaller than the missing energy carried by a neutralino
coming from the decay of a much heavier colored object.

Searches involving leptons usually require missing energy,
one or more leptons, and possibly some number of jets.
The latest ATLAS search of this type, for example, based on 
1~fb$^{-1}$ is described in~\cite{Aad:2011cw}. 
Slepton flavor mostly affects the two lepton channels,
since these are sensitive to two correlated leptons coming from the
decay chain 
\beq\label{decay}
\chi^0_2\to l^\pm \tilde l^\mp \to \chi^0_1 l^\pm l^\mp\,.
\eeq
Such searches were designed with the assumption
that the selectron and smuon are degenerate with no mixing,
so that the OS leptons of~\eqref{decay} are of the same flavor (SF).
This assumption is important in even the simplest counting-experiment
dilepton searches, since these rely on ``flavor subtraction'' 
in order to enhance the SUSY signal contribution over the SM background 
(and over the SUSY background of uncorrelated leptons), by measuring
\beq\label{fs}
N_{flav-sub}\equiv N(e^+ e^-) + N(\mu^+ \mu^-) - N(e^\pm \mu^\mp)\ .
\eeq
If the slepton spectrum is generation-independent, the supersymmetric
signal
contributes in only the first two terms of~\eqref{fs}, while the uncorrelated 
lepton SM background 
contributes equally to these and to the third term and therefore
drops out.
If the slepton spectrum is generation-dependent,  
$N_{flav-sub}$ might not be a sensible observable.
The relevant slepton states (between the heavier and lighter neutralinos)
are in general  
some combinations of the selectron, smuon and stau, with some mass
differences.
For simplicity, assume that two of the lightest sleptons
are selectron-smuon mixtures,
with  mixing $\sin\theta$,
\beqa\label{mixedslep}
\tilde l_1 &=& \cos\theta\, \tilde e -\sin\theta \,\tilde\mu\,,\nn\\
\tilde l_2 &=& \sin\theta \,\tilde e +\cos\theta \,\tilde\mu \,,
\eeqa
and with masses $m_1=m$ and $m_2=m+\delta m$,
such that 
\beq\label{ordering}
m_{\tilde\chi_1^0} < m < m+\delta m < m_{\tilde\chi_2^0}\,. 
\eeq
If the mixing is very small, $\sin\theta\ll1$,
flavor subtraction still works. Note however that
in this case, flavor constraints allow a substantial
slepton mass splitting $\delta m$, and if this splitting is not
much smaller than $m_{\chi^0_2}-m$,  the branching ratios
of $\chi^0_2$ to the different sleptons are different
and $N(e^+ e^-)\neq N(\mu^+ \mu^-)$.
If, on the other hand, the mixing is substantial,
the leptons in~\eqref{decay} can have either flavor.
In particular, both SF and DF dileptons appear in the decay~\eqref{decay},
and $N_{flav-sub}$ dilutes the signal.

Another common strategy (also employed in~\cite{Aad:2011cw},
see also~\cite{Eckel:2011pw})
for enhancing the SUSY signal over the background, is to measure
the OS dilepton invariant mass distribution. The reason is that
the signal distribution has  a triangle shape, which 
peaks at the kinematic endpoint of the dilepton
invariant mass~\cite{Hinchliffe:1996iu,Bachacou:1999zb,Gjelsten:2004ki,Lester:2005je,Lester:2006cf,Barr:2007hy,Autermann:2009js}, 
while the background from uncorrelated
leptons is roughly constant (at least over some range of the dilepton
invariant mass near the endpoint)\footnote{The dilepton invariant mass 
distribution is an important tool
for measuring superpartner masses and we will return to it
in sections~\ref{inverse} and~\ref{measure}.
For now, however, we are mainly interested in it at the level of
the discovery potential for SUSY.}.
If the slepton spectrum is generation-independent,
$\delta m=0$ and $\sin\theta=0$, and the dilepton-invariant
mass distribution has a single peak, at the endpoint of
$m_{ll}$, 
\begin{equation}\label{ep}
m_{ll}^2 | _{ep} = \frac{(m_{\tilde\chi_2^0}^2 - m_{\tilde l}^2)(m_{\tilde l}^2 
- m_{\tilde\chi_1^0}^2)}{m_{\tilde l}^2}\,.
\end{equation}
with identical contributions from the two
sleptons.
The resulting SF invariant mass distribution has a triangle shape
(see Figure~1).
\begin{figure}[ht]
\centering
\includegraphics[height=5.2cm]{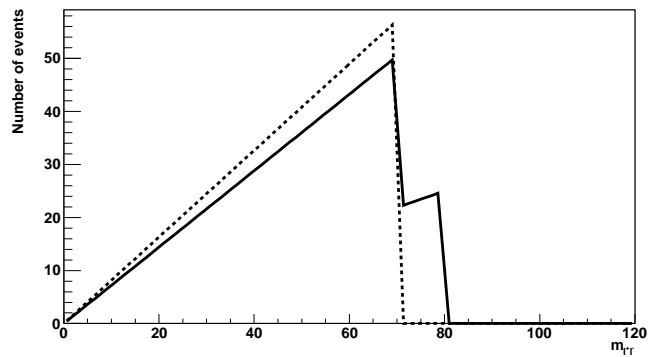}
\caption{The $l^+l^-$ invariant mass distribution ($l=e,\mu$) 
for degenerate sleptons (dashed) and for sleptons
of different masses (solid).} 
\label{Fig:triangle}
\end{figure}
With $\delta m\neq0$, the two sleptons give rise
to two separate endpoints, $m_{ll,1}$ and $m_{ll,2}$~\cite{Allanach:2008ib}. 
For small $\delta m$, the edge separation is
\begin{equation}\label{deltaedges}
\Delta m_{ll} \equiv m_{ll}| _2- m_{ll}| _1
\sim \,\frac{m_{ll}}{m_{\tilde l}}
\frac{m_{\chi_2^0}^2m_{\chi_1^0}^2-m_{\tilde l}^4}
{(m_{\chi_2^0}^2 - m_{\tilde l}^2)(m_{\tilde l}^2 - m_{\chi_1^0}^2)}
\, \delta m \,.
\end{equation}
Note that this edge separation can either enhance or suppress
the slepton mass difference.
In particular, for $m (m+\delta m)\sim  m_{\tilde\chi_1^0} m_{\tilde\chi_2^0}$
one finds
$\Delta m_{ll}\ll \delta m$, 
and for  $m (m+\delta m)$ very far from this value, 
$\Delta m_{ll}\gg \delta m$.

Consider first a small slepton mixing $\sin\theta\ll1$~\cite{Allanach:2008ib}.
In this case, the $ee$ distribution is only sensitive to the selectron,
and the $\mu\mu$ distribution is only sensitive to the smuon,
so flavor subtraction can be used.
The $m_{ll}$ distribution of SF dilepton pairs now exhibits
two endpoints, separated by $\delta m_{ll}$ (see Figure~1). 
If $\delta m_{ll}$
is large, the height of the corresponding peaks is smaller than
the height of the single peak in the generation-independent
model with $\delta m=0$ (see Figure~1).
It would therefore be harder to detect these peaks.

If, on the other hand,  $\delta m_{ll}$ is just somewhat above the possible
resolution of the experiment, the two endpoints
are one or two bins apart. Because of detector effects, the SF distribution 
will then exhibit a fuzzier peak, which again would be harder to observe
compared to the single peak of the flavor-blind spectrum.

A non-zero mixing complicates the picture, since now
the signal generates both SF and DF dileptons.
The two endpoints appear therefore in both the SF and DF 
$m_{ll}$ distribution, again with heights smaller than
the single peak obtained for $\delta m=0$.
As suggested in~\cite{Galon:2011wh}, a better approach in this case
would be  to consider the flavor-{\sl added} (ie, SF$+$DF) 
$m_{ll}$ distribution.
This distribution contains all the signal contributions to the two endpoints.
Furthermore, its shape  is independent of the mixing $\sin\theta$. 
Irrudicble backgrounds from
uncorrelated lepton pairs 
are expected to be roughly constant over reasonable ranges
of $m_{ll}$, so one should be able to extract them  by fitting this
distribution.

\section{Interpreting the new physics in the presence
of generation-dependence}\label{inverse}
If superpartners are eventually discovered, a lot of work
would be required in order to establish that they are indeed the
supersymmetric partners of SM particles 
(see, \eg~\cite{Feng:1995zd,ArkaniHamed:2005px}).
This would entail, beyond the obvious mass measurements,
also measuring the sizes of couplings of the new 
particles, and ultimately their spins.
The latter are particularly important since
different SM extensions
motivated by the hierarchy problem, most notably UED~\cite{Appelquist:2000nn}, 
have similar signatures at the LHC~\cite{Cheng:2002ab}. 
This problem has come to be known as
the ``Inverse Problem''~\cite{ArkaniHamed:2005px}, 
since it is the inverse of the traditional
approach to NP at colliders, whose starting point is a specific NP
model. 
Determining the spins of new particles would typically require
measuring angular distributions or other event shape variables.

Supersymmetric models and their look-alikes~\cite{Cheng:2002ab} often
have a stable neutral particle at the bottom of the spectrum,
with all other new particles decaying to this new lightest particle
 plus some SM particles.
As a result, NP events cannot be fully reconstructed,
and some sophisticated methods are required even for mass measurements.
The kinematic endpoint of  the dilepton invariant mass distribution
from~\eqref{decay} is a well-known tool for extracting superpartner
masses~\cite{Hinchliffe:1996iu,Bachacou:1999zb,Gjelsten:2004ki,Lester:2005je,Lester:2006cf,Barr:2007hy,Autermann:2009js}. 
Furthermore, since the $m_{ll}$ distribution depends
on the spin of the intermediate particle in the decay, it
has been extensively discussed as a useful discriminator between
different types of NP~\cite{Smillie:2005ar,Athanasiou:2006ef,Wang:2008sw}.
Both applications of the $m_{ll}$ distribution are  complicated 
by having a generation-dependent NP spectrum.
Here we focus on its use as a spin discriminator.

The sharp triangle form of the $m_{ll}$ distribution
for di-leptons from the SUSY decay~\eqref{decay} follows
from the fact that the intermediate slepton is a scalar.
In the slepton rest frame, $m_{ll}^2 \propto (1-\cos\theta)$ 
where $\cos\theta$ is the angle between the two leptons.
Since $d\sigma/dm_{ll}^2$ is constant, $d\sigma/dm_{ll}$
is a straight line. 
Different NP models give rise to analogs of the decay~\eqref{decay},
with the intermediate particle potentially having a nonzero
spin. In that case, the $m_{ll}$ distribution is no longer
a straight line. As shown in~\cite{Smillie:2005ar,Athanasiou:2006ef}, 
with a SUSY-like
mass spectrum, it would still be hard to differentiate 
between SUSY and UED models based on this distribution,
but other spin combinations can be distinguished.
Thus for example, replacing the intermediate slepton
by a vector particle results in a very different distribution.
For near-degenerate spectra, UED and SUSY can be better
differentiated. In any case, distinguishing between SUSY
and other frameworks relies on the triangular shape
of the SUSY distribution. 
As we saw in the previous section, in the presence of
flavor-dependence, the single triangle shape of the 
SUSY distribution is replaced by the  double triangle
of Figure~1.
This shape would be smeared by detector effects,
so that it would be harder to differentiate from 
the shapes obtained in other models.

\section{NP Flavor measurements: measuring the slepton mass
splittings and mixings}\label{measure}
If NP is discovered at the LHC, one would want to understand
its flavor structure\footnote{For early work on the collider
signatures of generation-dependent sleptons, focused on
lepton flavor violation, 
see, \eg,~\cite{ArkaniHamed:1996au,ArkaniHamed:1997km,Agashe:1999bm,Hisano:2002iy,Bartl:2005yy}.}.
This question is relevant even if the new particles do not
come in three copies corresponding to the three generations
of the SM.
For any new particle with couplings to the SM fermions,
one would like to know whether these couplings are generation
dependent. In the case of a chargino or neutralino,
this translates to information about the sfermion mixings.
At the same time, for any new particle discovered, one would
like to establish whether only a single particle is seen,
or a few nearly degenerate particles with similar signatures.
Thus for example, a supersymmetric model with a neutralino LSP
and generation blind slepton spectrum (with degenerate
selectrons and smuons), may be hard
to distinguish from a model with a single slepton
with equal selectron and smuon components.
Differentiating between the two involves information
on both the mass splittings and mixings.
It is important to note that nearly-degenerate particles
are a very plausible possibility, given the strong flavor
constraints on the first and second generation. 
As we saw in section~\ref{constraints}, these essentially restrict the product
of the mass splitting and mixing, so that a small mass
splitting is necessary if the mixing is non-negligible.

In the following we will discuss these questions in the 
context of the first and second generation sleptons,
focusing on mixed sleptons as in~\eqref{mixedslep}, with a small
mass splitting $\delta m$ of order a few GeV.
When discussing mass measurements, one must distinguish
between two qualitatively different scenarios.
One is supersymmetry with a neutralino LSP, leading
to missing energy in each event, and the second
is supersymmetry with a charged (N)LSP, such as 
a slepton, in which most events have no missing energy
and are therefore fully reconstructible.

\subsection{Neutralino LSP models}
As discussed in section~\ref{inverse}, 
the dilepton invariant mass distribution
is an important tool for mass  measurements in neutralino LSP models.
We have already seen that a slepton mass splitting
results in a double triangle shape of this distributions,
with the separation of the two endpoints being a non-trivial
function of the two neutralino masses, $m$ and $\delta m$\footnote{Here and
in the following we assume that the mass splitting is much bigger
than the slepton decay width. The effect of non-zero width was
studied in detail in~\cite{Grossman:2011nh}}.
This example was studied in detail in~\cite{Galon:2011wh}, and in the
following we summarize the results.
If the endpoint separation is large, it would be fairly
easy to detect the two endpoints, and to infer the presence of two distinct
slepton states of different masses.
Note, however, that a large endpoint separation is typically obtained
when the slepton masses are close to either one of the neutralino
masses [recall that when the two slepton masses are close to the 
geometric mean of the two neutralino masses, the edge separation 
$\Delta m_{ll}\ll \delta m$,  see~\eqref{deltaedges}]. 
In this case, one of the leptons emitted
in the decay~\eqref{decay} is relatively soft,
making the endpoint measurement more challenging~\cite{Buras:2009sg}. 
Indeed, for this reason, most benchmark points
chosen for measurements of the dilepton invariant mass distribution
had (degenerate) sleptons far from the two neutralinos.

As discussed in section~\ref{search}, in the presence of mixing,
the decay~\eqref{decay} contributes both SF and DF dileptons.
The flavor-{\sl added} distribution is then particularly useful,
since it does not dilute the signal. Another advantage
of the flavor-added distribution is that it is independent of the
mixing $\sin\theta$. 
One can therefore start by extracting the two endpoints
from the flavor-added distribution (by fitting it to
a double triangle), and then use these endpoints as input to
a simultaneous
fit of the $ee$, $\mu\mu$ and $e\mu$ distributions.
The total numbers of events in these distributions are related by,
\begin{eqnarray}
\label{eq:event_ratios}
\frac{N_{FV}(e^\pm \mu^\mp)}{N_{FV}(e^+e^-)} & = & 
\frac{2(1+R) \cos^2 \theta \sin^2 \theta}{\cos^4\theta + R\sin^4 \theta} \\ \nonumber
\frac{N_{FV}(\mu^+ \mu^-)}{N_{FV}(e^+e^-)} & = & \frac{R\cos^4\theta 
+ \sin^4 \theta}{\cos^4\theta + R\sin^4 \theta}\,,
\end{eqnarray}
where $R$ is the ratio of phase-space factors in $\chi_2^0$ decays involving 
the different intermediate sleptons:
\begin{equation}
R \equiv \left(\frac{m_{\chi_2^0}^2 - m_{\tilde l_2}^2}{m_{\chi_2^0}^2 
- m_{\tilde l_1}^2}\right)^2\,.
\end{equation} 
Fitting each one of the distributions with a double triangle with
the endpoints as input, one can extract the mixing angle.
In particular,
for $\delta m\ll m_{\chi_2^0} - m$, $R$ is approximately 1,
so the fit only depends on the mixing angle.
These methods were applied in~\cite{Galon:2011wh} to an example 
with $\delta m\sim 3$~GeV, with the endpoints separated by 6~GeV,
showing (at a 14~TeV LHC with 10~fb$^{-1}$)
that the two endpoints and the mixing can be resolved  
both for small mixing 
($\sin^2\theta\sim 0.03$) and for large mixing ($\sin^2\theta\sim 0.6$).

Note that $R$ and the two endpoints contain complementary information
about the slepton and neutralino masses.
As stressed above, a small endpoint difference $\Delta m_{ll}$ does
not necessarily mean a small $\delta m$. 
If $\Delta m_{ll}$ is small, 
$R\neq1$ would indicate an appreciable slepton mass splitting,
with the sleptons close to the mean of the neutralino masses.

\subsection{Slepton  NLSP models}
Models with a metastable NLSP charged slepton can be obtained
with gauge-mediated
supersymmetry breaking~\cite{Feng:1997zr}, and in large regions
of the parameter space of gravity-mediated supersymmetry
breaking~\cite{Feng:2003xh}.
The NLSP  eventually decays to a gravitino, but if this decay occurs
outside the detector, the slepton leaves a
track in the muon detector, its momentum is measured, 
and its mass can be determined based on its 
time-of-flight or energy deposition patterns~\cite{Nisati:1997gb,Connolly:1999dv,Tarem:2009zz,Aad:2011hz,Khachatryan:2011ts,Chen:2009gu}\footnote{Apart from~\cite{Chen:2009gu}, these techniques rely on the low
slepton speed. Sleptons produced in the decay of heavier particles
can have large boosts, and therefore look like fake muons. 
Such sleptons may be detected by looking for a peak in the muon-lepton
invariant mass distribution~\cite{Galon:2011ws}.}.
Supersymmetric events are then not only fully reconstructible,
but also virtually background-free.

Measurements of masses and mixings in such models were discussed
in~\cite{Feng:2007ke,Kitano:2008en,Kaneko:2008re,Feng:2009yq,Feng:2009bd,Ito:2009xy}. 
A clean way of extracting the masses is
to sequentially reconstruct each particle in a cascade decay
by looking for the peak it generates~\cite{Feng:2009bd}.
Consider for example the decay of a heavy slepton
\beq
\tilde l_{{\rm heavy}} \to \chi_1^0 l_a  \to 
\tilde l_{{\rm NLSP}} l_b l_a \,,
\eeq
where $\tilde l_{{\rm NLSP}}$ is the metastable slepton.
The $\chi_1^0$ can be identified through the peak in the
OS  lepton-$\tilde l_{{\rm NLSP}}$ invariant mass distribution.
The next superpartner in the chain,   $\tilde l_{{\rm heavy}}$,
is often hard to detect by looking for a peak in the  
 $ll\tilde l_{{\rm NLSP}}$ 
distribution, due to the large combinatorial background.
Choosing however only those $l\tilde l_{{\rm NLSP}}$ pairs whose
invariant mass is close to the $\chi_1^0$ peak, 
one can construct a sample of $\chi_1^0$ candidates,
and then add to each of them another lepton. 
The resulting lepton-$\chi_1^0$-candidate
invariant mass distribution would exhibit a clearer $\tilde l_{{\rm heavy}}$ peak.

Once the masses are determined, the mixings can be measured
by comparing the relative numbers of events contributing
to peaks involving different flavor leptons. Thus for example,
comparing the number of events in the $\chi_1^0$ peak of
the $e\tilde l_{{\rm NLSP}}$, $\mu\tilde l_{{\rm NLSP}}$, and  
$\tau\tilde l_{{\rm NLSP}}$,
resolves the flavor composition of the $\tilde l_{{\rm NLSP}}$,
and similarly for the heavier sleptons.

As explained above, a particularly plausible scenario
is that two sleptons are nearly degenerate. If both of these
sleptons have observable decays, as would
be the case for example for nearly degenerate $\tilde l_{{\rm heavy}}$'s,
one would see two different peaks in the $l\chi_1^0$ invariant-mass
distributions.
Whether or not the  two peaks can be resolved depends 
both on the $\tilde l_{{\rm heavy}}$ masses and on their flavor composition.
An even more challenging scenario however is that two (or more)
of the lighter sleptons are nearly degenerate.
Consider for concreteness a spectrum with a second slepton,
$\tilde l_2$, a few GeV above the mass of $\tilde l_{{\rm NLSP}}$,
so that $\tilde l_2$ decays to $\tilde l_{{\rm NLSP}}$ via a three-body 
decay  $\tilde l_2  \to \tilde l_{{\rm NLSP}} X$, where
$X$ is a dilepton~\cite{Ambrosanio:1997bq,Feng:2009bs}. 
The two leptons in $X$
are typically soft because of the small slepton mass difference,
and will go undetected, so that the $\tilde l_2$ decays
are invisible.
The existence of $\tilde l_2$ can then be probed using the
``Shifted Peak'' method, proposed in~\cite{Feng:2009yq},
which utilizes the hard lepton $l_2$ emitted in association with $\tilde l_2$.
The neutralino $\chi_1^0$
has two possible decays into sleptons.
The first is the direct decay to $\slep_{{\rm NLSP}}$,
\begin{equation}\label{ee}
\chi_1^0\to\tilde l_{{\rm NLSP}}^\pm l_1^\mp \ .
\end{equation}
The second is the decay to $\tilde l_2$,
\begin{equation}\label{mm}
\chi_1^0\to \slep_2^\pm l_2^\mp \ ,
\end{equation}
followed by one of the two three-body
decays~\cite{Ambrosanio:1997bq,Feng:2009bs}
\begin{eqnarray}\label{xz}
\slep_2^\pm&\to& \slep_{{\rm NLSP}}^\pm X^{\pm \mp} \ , \\
\slep_2^\pm&\to& \slep_{{\rm NLSP}}^\mp X^{\pm\pm}\ , \label{xc}
\end{eqnarray}
where $X^{\pm \mp}$ contains two OS leptons, and
$X^{\pm\pm}$ contains two SS leptons.  Note that the
charge-flipping decays of Eq.~(\ref{xc}) resulting in SS leptons
are possible because the neutralino is a Majorana fermion. 
SS leptons
are also present in models other than supersymmetry when the decay is
mediated by a vector boson or a scalar. 
Thus, the observed particles are the hard lepton from~\eqref{ee} 
or \eqref{mm}, and the long-lived slepton
$\slep_{{\rm NLSP}}$ from \eqref{ee}, \eqref{xz}, or \eqref{xc}.

We can thus construct distributions for the following invariant
masses-squared:
\begin{eqnarray}
m_{\slep l_1}^2&\equiv& \left(p_{\slep_{{\rm NLSP}}} +p_{l_1}\right)^2 \ ,\label{slep1}\\
m_{\slep l_2}^2&\equiv& \left(p_{\slep_{{\rm NLSP}} } +p_{l_2}\right)^2 \ ,\label{slep2}
\end{eqnarray}
where the $\slep_{{\rm NLSP}}$ and $l_2$ charges can be either opposite or the
same.  
The distribution~\eqref{slep1} peaks of course at the neutralino mass
$m_{\chi_1^0}$, but, because of the missing leptons,~\eqref{slep2}  
peaks somewhat below this mass,
at $m_{\chi_1^0}-\ES$,
with
\begin{equation}\label{shift}
\ES = \frac{m_{\chi_1^0}^2+m^2}{2 m m_{\chi_1^0} }\, \delta{m}\,.
\end{equation}
The slepton mass difference $\delta m$ can therefore be extracted
from the shift $\ES$ between the two peak locations.

The charge-flipping decays of~\eqref{xc} provide a useful
handle on the shifted peak.
 A peak in the invariant mass $m^2_{\slep_{{\rm NLSP}}^\pm l_2^\pm}$,
formed from events with SS sleptons and leptons, can only come from
the decays of~\eqref{mm} and will therefore exhibit the shift
$\ES$. The analogous OS distribution will contain both types of events
specified in~\eqref{ee} and~\eqref{mm}, and  will therefore
generically exhibit
a double peak structure, with the two peaks separated by $\ES$.

The identities of $l_1$ and $l_2$ depend, of course, on the flavor
compositions of $\slep_{{\rm NLSP}}$ and $\slep_2$.  In one
extreme case, if these are the left- and right-handed sleptons
associated with the same flavor, the two leptons are identical.
In the opposite extreme, the two sleptons could be pure states of
different flavors.  In this case, the leptons $l_1$ and $l_2$ are
different flavors, and there is no need to rely on the charges to
separate the distributions.  
More generally, the two sleptons contain both selectron and smuon components,
and each of the OS $\tilde l_{{\rm NLSP}}e$ and $\tilde l_{{\rm NLSP}}\mu$ 
distributions will exhibit both the true neutralino peak and the shifted
peak. Still, the SS distributions will be sensitive to just the $\tilde l_2$,
and can therefore be used to cleanly determine its flavor composition.
In~\cite{Feng:2009bd}, it was shown that mixings of around 5$\%$ and
mass differences of around 5~GeV can be resolved in this way.

\section{Acknowledgments}
Research supported in part by the Israel Science
Foundation (ISF) under grant No.~1367/11, by the United States-Israel
Binational Science Foundation (BSF) under grant No.~2010221.

\end{document}